\documentclass[10pt,letterpaper]{article}
\usepackage{opex3,amsmath,amssymb}

\newcommand{\ket}[1]{\left | \, #1 \right \rangle}
\newcommand{\bra}[1]{\left \langle #1 \, \right |}

\newcommand{\mi}{\mathrm{i}}
\newcommand{\rmd}{\mathrm{d}}

\begin{document}

\title{Large-Alphabet Time-Frequency Entangled Quantum Key Distribution by means of Time-to-Frequency Conversion}

\author{J.~Nunn$^1$, L.~Wright$^1$, C.~S\"{o}ller$^{1,2}$, L.~Zhang$^3$, I.~A.~Walmsley$^1$ and B.~J.~Smith$^{1*}$}
\address{$^1$Clarendon Laboratory, University of Oxford, Parks Road, Oxford OX1 3PU, UK}
\address{$^2$Current address: Heraeus Noblelight GmbH, Heraeusstra\ss e 12-14, 63450 Hanau, Germany}
\address{$^3$Bldg.~99, Max Planck Research Department for Structural Dynamics, University of Hamburg, Luruper Chaussee 149, 22761 Hamburg, Germany}
\email{$^*$b.smith1@physics.ox.ac.uk} 



\begin{abstract}
We introduce a novel time-frequency quantum key distribution (TFQKD) scheme based on photon pairs entangled in these two conjugate degrees of freedom. The scheme uses spectral detection and phase modulation to enable measurements in the temporal basis by means of time-to-frequency conversion. This allows large-alphabet encoding to be implemented with realistic components. A general security analysis for TFQKD with binned measurements reveals a close connection with finite-dimensional QKD protocols and enables analysis of the effects of dark counts on the secure key size.
\end{abstract}

\ocis{270.5565, 060.5565 (Quantum communications); 270.5568 (Quantum cryptography); 270.5585 (Quantum information and processing); 060.4230 (Multiplexing); 040.5570, 270.5570 (Quantum detectors); 060.4080 (Modulation); 060.5060 (Phase modulation); 060.4510 (Optical communications).} 

\bibliography{Physics_combi_tfqkd}
\bibliographystyle{unsrt}
%
%

\section{Introduction}
Light and matter are ultimately constituted of quantum entities, so that the limitations of technology cannot be prescribed by the principles of classical physics. Quantum-optical devices promise better-than-classical performance in sensing \cite{Rarity:1990fk,Datta2011}, computation \cite{shor1997pta,lanyon:250505,Martin-Lopez:2012vn} and, in particular, communications \cite{Muller:1997kx,Shapiro:2009ys}. Quantum effects in optical systems are easily observed at room temperature because of the high carrier frequency of optical signals compared with thermal background radiation. Light has numerous degrees of freedom that can be co-opted for the transmission of quantum information, and optical signals can be generated and detected efficiently. Modern telecommunications leverage the very large capacity of optical channels via dense multiplexing techniques, and the efficiency of quantum information processing can be similarly enhanced with the use of large-alphabet encoding strategies. In this connection, time-frequency encoding offers numerous advantages over polarisation or spatial-mode encodings, since multiple spectral modes are supported in waveguides and optical fibres with minimal dispersion, and zero cross-talk. For this reason, large-alphabet spectral-temporal multiplexing in quantum optics has recently received growing attention \cite{Hayat:2012kx,Rohde:2012ve}, with immediate applications in \emph{quantum key distribution} (QKD)\cite{Mower:2012cr,Chang-Hua:2010qf,Qi2006}.

Public key cryptography underpins much of commerce today \cite{Galbraith:2012fk}, but it is threatened by the advanced computational tools made possible by quantum information processing. Although a quantum computer remains many years away, the threat is sufficient to motivate the search for security that is immune to a quantum adversary. Quantum key distribution \cite{Gisin:2002xe} provides this security, and in the three decades since the first proposals were made based on polarized photons \cite{bennett1984qcp}, it has developed into both a burgeoning industry \cite{Industries}, and a diverse research field \cite{Simon:2012uq,Gauthier:2012kx,Braunstein:2012vn,Lo:2012ys}. While polarisation states are vulnerable to corruption in waveguides and optical fibres, the arrival time of photons is a more robust encoding \cite{gisin-2004}. In general, transmission losses limit the range of QKD, but the use of entangled photons \cite{Ekert1991} allows the distance to be extended by means of quantum repeaters \cite{Duan:2001zr, Saglamyurek:2011ly}, which employ entanglement swapping between short segments to distribute secure correlations over large distances. 

These considerations motivate the study of QKD with photon pairs entangled in their chronocyclic degree of freedom \cite{Qi:2011bh,Mower2011,Takesue:2010dq, Ali-Khan2007,Olislager2010,Hayat:2012kx}. The security of such time-frequency QKD (TFQKD) protocols is based on the ability to detect non-classical correlations in at least two complementary measurement bases, and these bases should be mutually unbiased (\emph{i.e.} Fourier conjugates of one another) to maximise the sensitivity of the measured correlations to intrusion by an eavesdropper. Time and frequency domain detection provides a particularly convenient conjugate pair both because time- and frequency-resolving measurements are mature technologies, and because photon sources based on parametric scattering, such as spontaneous downconversion or four-wave mixing naturally produce photon pairs with widely-tunable correlation properties ranging from uncorrelated emission \cite{Mosley2008,Cohen2009} to extremely broadband spectral entanglement \cite{Brida_2009,Grice1997,zhang2007ghe}. However, accessing the large entropy of these photon pairs requires precise measurements in both time and frequency bases, which is a non-trivial problem because of technical constraints on the timing jitter of photon detectors and the resolution of spectrometers.

Recently it was proposed to use fast time-resolving photon detectors to record the correlated arrival times of entangled photons, and to insert a dispersive medium before the detectors to measure in the frequency basis \cite{Mower:2012cr}. The dispersion delays each spectral component of the incident photons differently, converting the arrival time at the detector into a frequency measurement \cite{Avenhaus:2009nx}. This protocol has the advantage that a single detector can sample a large region in chronocyclic space \cite{Walmsley:2009oq}. However the timing jitter of the detector sets a lower limit to the correlation time of the photon pairs. Fast detectors are not yet efficient \cite{Takesue:2007kl,cova198920}, whereas efficient detectors are generally slower \cite{Lita:2008tg,Hadfield2009,Marsili:2013bh}, so with current technology the spectral correlations should be very narrowband in order for the arrival-time correlations to extend over a long enough period to be resolved by the detectors. Besides the challenges of engineering a bright photon source to fit these criteria, the narrow bandwidth also means that extremely dispersive media are required to implement the frequency measurements, which are generally lossy. It was found in \cite{Mower:2012cr} that a photon efficiency of up to 2 bits per photon could be achieved with realistic components.

Here we introduce an alternative large-alphabet entangled TFQKD protocol that is dual to \cite{Mower:2012cr}. We propose to use spectral measurements to detect frequency correlations, and to use phase modulators to convert these measurements to the time basis. We show that with this detection scheme it is feasible to demultiplex up to 4 bits per photon with readily available components, while technical advances --- essentially improving the power dissipation of electro-optical phase modulators to allow larger modulation depth --- could allow much larger alphabets. While chronocyclic entanglement produces correlations between the continuous variables of arrival time and frequency, measurements inevitably map these correlations to a set of discrete outcomes. We provide a security analysis that accounts for the effects of this detector-binning, and we model the constraints on the secure key arising from dark counts.
\section{The protocol}
\begin{figure}[h]
\begin{center}
\includegraphics[width = 8cm]{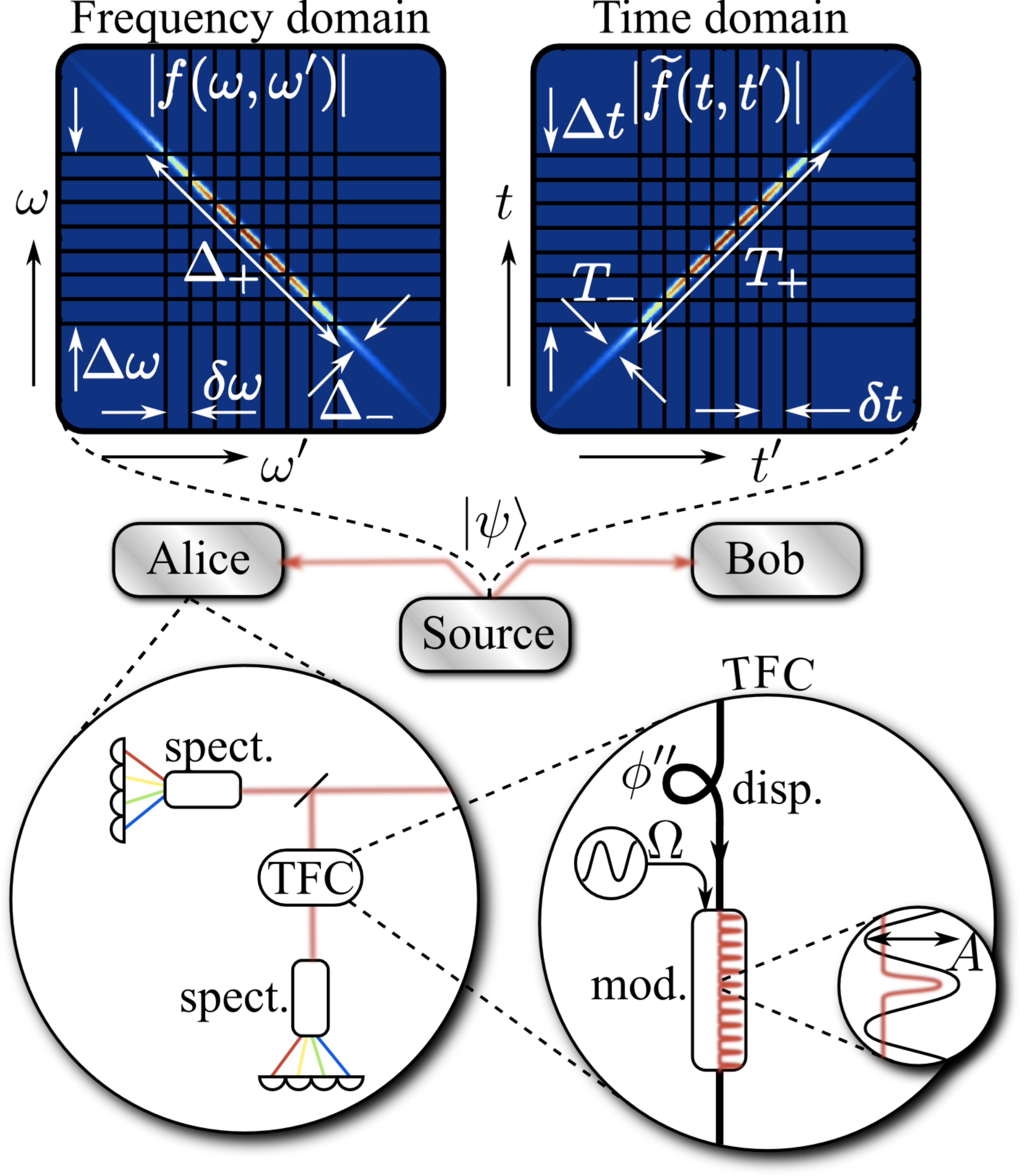}
\caption{Time-frequency quantum key distribution with spectrally entangled photon pairs. Top: we assume a Gaussian anti-correlated joint spectral amplitude $f(\omega,\omega')$ (left) and a corresponding positively correlated Gaussian arrival time distribution $\widetilde{f}(t,t')$ (right). Black ruled lines indicate spectral/temporal measurements made by Alice and Bob centred at the values $\{\omega_j\}/\{t_j\}$ (see text). Middle: photon pairs from the source are distributed to Alice and Bob, who have identical measurement setups. Bottom: spectral correlations are revealed with photon-counting spectrometers (spect). Security is certified by measurements in the conjugate arrival-time basis using time-to-frequency conversion (TFC), which is implemented with a dispersive element (disp) followed by a phase modulator (mod).}
\label{fig:TF_QKD_setup}
\end{center}
\end{figure}

Key distribution aims at generating a pair of correlated random bit strings shared exclusively by two parties --- conventionally called \emph{Alice} and \emph{Bob} --- that can be used to encrypt and decrypt communications by means of a \emph{one time pad} \cite{Vernam:1926bs}. In our TFQKD protocol, Alice and Bob construct correlated bit strings by recording the results of joint measurements on pairs of spectrally entangled photons. The entanglement produces correlations in both the frequencies and the arrival times of the photons which are destroyed if the system is measured by a third party. Uncorrelated measurement results therefore reveal the presence of an eavesdropper (\emph{Eve}) when Alice and Bob compare a subset of their bits. As shown in Fig.~\ref{fig:TF_QKD_setup}, a parametric source (which may or may not be under the control of Eve) produces photon pairs in the entangled state (ignoring for the moment the emission of vacuum or double pairs)
\begin{equation}
\label{psi}
\ket{\!\psi} = \int \!\!\! \int f(\omega,\omega') \ket{\omega,\omega'}\,\rmd \omega\,\rmd\omega',
\end{equation}
where $\ket{\omega,\omega'}$ indicates a photon pair received by Alice and Bob with frequencies $\omega+\omega_0$ and $\omega'+\omega_0$, respectively, with $\omega_0$ the central frequency of the pair. Here $f(\omega,\omega')$ is the joint spectral amplitude describing the frequency correlations between a photon pair emitted with detunings $\omega$, $\omega'$ from degeneracy. In general, energy conservation and phasematching considerations produce non-factorable joint spectra, $f(\omega,\omega')\neq f_\mathrm{A}(\omega)f_\mathrm{B}(\omega')$ \cite{Grice1998,Law2000}, and the degree of entanglement --- the number of correlated spectral modes --- is conveniently quantified by the \emph{Schmidt number} $K=1/\sum_j \lambda_j^4$, where the $\{\lambda_j\}$ are the singular values of $f$. Broad spectral entanglement with $K\sim 10^3$ has been observed \cite{Brida:2009ij} and routes to extremely multimode emission with $K\sim 10^7$ have been proposed \cite{zhang2007ghe}. Therefore although source engineering is non-trivial, it is already possible to produce photon pairs with a high degree of quantum correlations. As discussed above, the primary challenge for implementing TFQKD lies with the detection scheme. In our protocol, Alice and Bob each have a photon-counting spectrometer, which they use to measure the frequencies of their photons. A photon entering one of the spectrometers is spatially dispersed and then directed towards one of $M$ photon counting detectors, such that a `click' at the $j^\mathrm{th}$ detector ($j=1,2,\ldots,M$) is described by a projective positive operator-valued measure (POVM) element
\begin{equation}
\label{POVM1}
\Pi_j = \int F_j(\omega)\ket{\omega}\bra{\omega}\,\rmd \omega.
\end{equation}
Here $F_j$ is the instrument response that represents the finite resolution of the spectrometer --- that is, the range of frequencies coupled to the $j^\mathrm{th}$ detector. For simplicity in the calculations to follow we will assume a `top hat' profile
\begin{equation}
\label{F_bin}
F_j(\omega) =\left\{\begin{array}{cc}1\quad & | \omega-\omega_j | \leq \delta\omega/2;\\
0 \quad& \quad\quad\quad\;\mathrm{otherwise},\end{array}\right.
\end{equation}
with $\omega_j = \omega_0 + [j-(M+1)/2]\delta\omega$. This describes a spectrometer with a resolution $\delta \omega$ and a flat response to all frequencies in the range $\omega_0 -\Delta\omega / 2 \leq \omega \leq \omega_0 +\Delta\omega/2$, where $\Delta \omega = M\delta \omega$. This description would be appropriate for a grating spectrometer equipped either with a photon counting charge-coupled device (CCD), or a v-groove array of fibres coupled to Geiger-mode avalanche photodiodes (APDs), for example \cite{Zhang:2009uq}. The entanglement of the state in Eq.~(\ref{psi}) will cause correlated clicks at the outputs of Alice's and Bob's spectrometers, which they can use to generate a cryptographic key. The frequency measurements have $M$ possible outcomes, so that a detection event can extract up to a maximum of $I_M=\log_2 M$ bits of information from each photon pair.

 However to certify the security of their key, they need to also check that the arrival times of the photons are correlated. As mentioned previously, the large timing jitter ($\gtrsim$200~ps for APDs) of efficient photon-counting detectors makes it challenging to resolve the temporal correlations directly. Instead, we propose to implement measurements in the time basis by sending each photon through a \emph{time-to-frequency converter} \cite{Kolner:1994fv,Bennett1999,LavoieJ.:2013ly}. As shown in Fig.~\ref{fig:TF_QKD_setup}, this comprises a dispersive element followed by a phase modulator, the combined effect of which is to implement a Fourier transform on the incident waveform, thus converting time to frequency. The effect is rigorously analogous to a spatial Fourier transform implemented on the transverse profile of a light field using a lens followed by free-propagation over a distance equal to its focal length. To see how this works, consider first the effect of the phase modulator on an incoming field (analogous to free propagation alone). The modulator imprints a sinusoidally time-varying\footnote{Other phase modulation schemes such as four-wave mixing in waveguides \cite{Salem:2008uq} or sum-frequency generation in nonlinear crystals \cite{Bennett1999} could of course be used. Here we consider electro-optic phase modulation, which can be implemented with off-the-shelf telecoms components efficiently and with low loss.} phase $\phi(t)=A\cos(\Omega t)$ with amplitude $A$ and angular frequency $\Omega$. Near the peak of the modulation at $t\approx 0$ the phase has a quadratic time dependence $\phi(t) \approx -A\Omega^2t^2/2 = -\ddot{\phi}t^2/2$, where we have defined $\ddot{\phi} = A\Omega^2$ and we dropped an unimportant global phase. The time-dependent amplitude $E(t)$ of the incident field is transformed to $E(t)\exp\{\mi \ddot{\phi}t^2/2\}$, which in frequency space induces a transformation analogous to the Fresnel-Kirchhoff diffraction formula,
\begin{equation}
\label{Mod1}
\ket{\!\omega} \longrightarrow (2\pi \mi \ddot{\phi})^{-1/2} \int e^{\mi (\omega-\omega')^2/2\ddot{\phi}} \ket{\!\omega'}\,\rmd \omega'.
\end{equation}
We now consider that the field has previously propagated through a system with quadratic, or \emph{group velocity} dispersion, which imparts a spectral phase $\varphi(\omega) = -\varphi'' \omega^2/2$ (analogous to a lens). Inserting this phase and multiplying out the Gaussian exponent in Eq.~(\ref{Mod1}), the full transformation becomes
\begin{equation}
\label{Mod2}
\ket{\omega} \longrightarrow (2\pi \mi \ddot{\phi})^{-1/2} e^{\mi \omega^2/2\ddot{\phi}} \int e^{\mi(1/\ddot{\phi}-\varphi'')\omega'^2/2} e^{-\mi \omega \omega'/\ddot{\phi}} \ket{\omega'}\,\rmd \omega'.
\end{equation}
The choice $\varphi'' = 1/\ddot{\phi}$ corresponds to the \emph{spectral Fraunhofer limit} \cite{Azana:2003fk} in which the quadratic phase factor in the integrand of Eq.~(\ref{Mod2}) is removed, and the transformation is then operationally identical to a Fourier transform (the analogous cancellation of Fresnel diffraction in the focal plane of a lens is well-known). That is, any measurement of the spectral intensity of the transformed field is equivalent to a measurement of the temporal intensity --- the \emph{arrival time} --- of the input field, described by the POVM element
\begin{equation}
\label{POVM2}
\widetilde{\Pi}_j = (2\pi\ddot{\phi})^{-1/2}\int\!\!\!\int \widetilde{F}_j(\omega-\omega')\ket{\omega}\bra{\omega'}\,\rmd \omega\,\rmd \omega',
\end{equation}
where we have defined
\begin{eqnarray}
\label{FT}
\widetilde{F}_j(x) &=& (2\pi \ddot{\phi})^{-1/2}\int F_j(x')e^{-\mi x x'/\ddot{\phi}}\,\rmd x'\\
 &=& \delta\omega  (2\pi \ddot{\phi})^{-1/2} e^{-\mi \omega_j x/\ddot{\phi}}\mathrm{sinc}(x\delta \omega/2\ddot{\phi}).
\end{eqnarray}
The temporal resolution of this measurement is $\delta t = \delta \omega /\ddot{\phi}$, where $\delta \omega$ is the spectral resolution of the photon-counting spectrometer, as described above. Note that this is independent of the timing jitter of the photon counters used to register the photons. Like an ordinary spatial lens, the time lens has a finite aperture $\tau$, which is the region in time over which the phase modulation remains quadratic. To a good approximation this is given by $\tau = 1/\Omega$ \cite{Kolner:1994fv}. This sets a limit to how fast the modulator can be run, since the signals to be detected must pass through the aperture within this time window.

Having described how to implement temporal measurements, we consider the parameters required for an effective TFQKD protocol using our approach. As an idealised example, we assume a Gaussian correlated joint spectrum of the form \cite{Wasilewski2006,Grice2001}
\begin{equation}
\label{JSA}
f(\omega,\omega') = (\pi\Delta_+ \Delta_-/8)^{-1/2} \exp\left({-\omega_-^2/2\Delta_+^2 -\omega_+^2/2\Delta_-^2}\right),
\end{equation}
where $\omega_\pm = (\omega\pm \omega')/\sqrt{2}$ and $\Delta_+$, $\Delta_-$ are the marginal and correlation bandwidths, respectively. The Schmidt number is given by $2K = \Delta_+/ \Delta_- + \Delta_-/ \Delta_+$, which is roughly the ratio of the major and minor widths of the joint spectrum (see Fig.~\ref{fig:TF_QKD_setup}). This coincides with the intuition that the number of correlated modes is found by dividing up the joint spectrum and counting the number of bins of width roughly equal to the correlation bandwidth $\Delta_-$. To access these correlated modes with our detectors, we engineer the source such that $\Delta_- = \beta_- \delta \omega$ and $\Delta_+ = \beta_+ \Delta\omega$, where the constants $\beta_\pm$ are chosen so that the spectral resolution roughly matches the correlation bandwidth, and the spectral response covers the full width of the joint spectrum. For example, the choice $\beta_+=3/4$ yields a reasonably flat marginal probability distribution across all detection channels, while choosing $\beta_-=1/5$ provides sufficiently tight correlations that cross-talk between channels is negligible.

The correlations in the time domain are determined by the joint temporal amplitude $\widetilde{f}(t,t')$, which is the two-dimensional Fourier transform of the joint spectral amplitude, $f(\omega,\omega')$. The form assumed in Eq.~(\ref{JSA}) is convenient because $\widetilde{f}(t,t')$ is again a correlated Gaussian as shown in Fig.~\ref{fig:TF_QKD_setup}, with major and minor temporal widths given by $T_\pm = 1/\Delta_\mp$. If we require the same ability to resolve temporal correlations as spectral correlations, we should have a temporal resolution $\delta t = T_-/\beta_-$, which fixes the modulator phase curvature  to be
\begin{eqnarray}
\nonumber
\ddot{\phi} &=& \frac{\delta \omega}{\delta t}\\
\nonumber &=& \Delta_-\Delta_+\\
\label{phi_fix} &=& \beta_+ \beta_- M\delta \omega^2.
\end{eqnarray}
We also require that the temporal aperture of the modulator is long enough to accommodate the full signal, $\tau\geq T_+$, which limits the modulator frequency to
\begin{equation}
\label{freq_lim}
\Omega \leq \beta_- \delta \omega.
\end{equation}
Since the modulation frequency is restricted, the modulation depth $A$ must be large enough to achieve the required phase curvature. The demands on the modulation depth are minimised by matching the modulator frequency and the spectrometer resolution to achieve equality in the above condition. This can be achieved with modern GHz-bandwidth modulators and a relatively modest grating spectrometer. For instance, using the choices for $\beta_\pm$ introduced above, a 50~GHz modulator could be used with a 2~nm-resolution spectrometer and telecoms-band photons. Operating at this limit, the modulation depth required to implement the time lens is
\begin{equation}
\label{mod_depth}
A = \frac{\beta_+}{\beta_-} M.
\end{equation}
This is a relatively demanding condition: the modulation depth must rise in proportion to the number of frequency bins used for the QKD protocol. That is, exponentially in the number of bits carried by each photon pair. However modulation depths of order $20 \pi$ radians are feasible with lithium-niobate modulators \cite{Kauffman1994}, which would allow $M=16$ (an alphabet size of $I_M=4$ bits per photon pair). We note finally that the dispersion required to reach the Fraunhofer limit in this example would be provided by $\sim$220~m of ordinary silica fibre, with a GVD on the order of 300~fs$^2$~cm$^{-1}$. Such a short length of fibre would introduce very small losses.
\section{Security}
The output of TFQKD is a digital bit string, as for all QKD protocols, but this is extracted from measurements on continuous degrees of freedom. Therefore it is not immediately obvious whether a security analysis developed for discrete-variable QKD protocols can be applied. To examine the security of TFQKD with entangled photons, we adapt arguments developed for spatially-encoded QKD \cite{Zhang2008} to the case of finite detector resolution \cite{Rudnicki:2012vn, Schneeloch2013, Tasca2012, Ray:2013uq}. To proceed, we consider the correlations revealed by the sifted joint probability distribution $\{p^\mathrm{BA}_{ba}\}$ describing the probability of Bob's $b^\mathrm{th}$ detector firing along with Alice's $a^\mathrm{th}$ detector, when both measure in the frequency basis. These correlations are quantified by the \emph{mutual information} $I_{BA}$,
\begin{equation}
\label{mutual1}
I_\mathrm{BA} = \sum_{ba} p^\mathrm{BA}_{ba} \log_2 \left(\frac{p^\mathrm{BA}_{ba}}{p^\mathrm{B}_b p^\mathrm{A}_a}\right),
\end{equation}
where $p^\mathrm{B}_b = \sum_a p^\mathrm{BA}_{ba}$ is the marginal distribution of Bob's outcomes, and where $p^\mathrm{A}_a$ is similarly Alice's marginal distribution. For simplicity we suppose that Alice and Bob post-process their sifted key bits using a one-way forward reconciliation protocol, for which the size --- in bits --- of the secret key they can distill is bounded by the excess information
$$
I = I_\mathrm{BA}-I_\mathrm{BE},
$$
where $I_{BE}$ is the mutual information between Bob and Eve\footnote{In general the key distillation will have non-unit efficiency $\beta$ such that $I=\beta I_\mathrm{BA}-I_\mathrm{BE}$. However high efficiencies can be achieved, with $\beta > 90\%$, so that this represents a small correction. In any case, although the formulas are simplified by assuming $\beta=1$, our security analysis can be trivially extended to cover the general case.}. We can calculate $I_\mathrm{BA}$ from the statistics of Alice's and Bob's measurements but $I_\mathrm{BE}$ is not known and must be bounded by analyzing the correlations in the complementary (temporal) basis. To proceed, we write the mutual information in the form $I_\mathrm{XY} = H_\mathrm{X}-H_{\mathrm{X|Y}}$ \cite{Zhang2008}, where $H_\mathrm{X} = -\sum_x p^\mathrm{X}_x\log_2 p^\mathrm{X}_x$ is the marginal entropy and where the conditional entropy is given by
\begin{equation}
\label{entropy1}
H_\mathrm{X|Y}=\sum_y p^\mathrm{Y}_y H_{\mathrm{X}|\mathrm{Y}=y},\qquad \mathrm{where}\quad H_{\mathrm{X}|\mathrm{Y}=y}=-\sum_x p^\mathrm{XY}_{x|y}\log_2 p^\mathrm{XY}_{x|y}
\end{equation}
is the entropy associated with the conditional probabilities $p^\mathrm{XY}_{x|y}$ that the $x^\mathrm{th}$ detector of party X fires, given that the $y^\mathrm{th}$ detector of party Y fires. We then obtain
\begin{equation}
\label{secret}
I = H_\mathrm{B|E} - H_\mathrm{B|A}.
\end{equation}
Next we need to find a lower bound for $H_\mathrm{B|E}$. To do this, we note that the variability of Bob's measurement results in time and frequency are restricted by Heisenberg's uncertainty principle. More generally, if Bob receives an arbitrary quantum state $\rho$, the marginal entropies $H_\mathrm{B}(\rho)$ ($\widetilde{H}_\mathrm{B}(\rho)$) for Bob's measurements in frequency (time), satisfy the entropic uncertainty relation
\begin{equation}
\label{entropic_bound1}
H_\mathrm{B}(\rho) + \widetilde{H}_\mathrm{B}(\rho) \geq B,
\end{equation}
where $B$ is a number that depends only on Bob's measurements, not on the state $\rho$. Alice's and Eve's measurements on the state emitted by the source produce a conditional state $\rho_{\mathrm{B}|\mathrm{A}=a,\mathrm{E}=e}$ at Bob's detectors, so that we have $H_\mathrm{B}(\rho_{\mathrm{B}|\mathrm{A}=a,\mathrm{E}=e}) = H_{\mathrm{B}|\mathrm{A}=a,\mathrm{E}=e}$, using a natural tripartite extension of the notation in Eq.~(\ref{entropy1}). Substituting this conditional state into Eq.~(\ref{entropic_bound1}) and averaging over the possible outcomes of Alice's and Eve's measurements, we obtain
\begin{equation}
\label{bound2}
H_\mathrm{B|EA} + \widetilde{H}_\mathrm{B|EA} \geq B.
\end{equation}
Now, since uncertainty (entropy) is only increased by removing a ``given'' precondition from a conditional distribution, we can remove Alice as a given from the first conditional entropy, and Eve as a given from the second, to get
\begin{equation}
\label{bound3}
H_\mathrm{B|E} + \widetilde{H}_\mathrm{B|A} \geq B.
\end{equation}
Inserting this into Eq.~(\ref{secret}) gives
\begin{equation}
\label{I_bound}
I \geq B-\widetilde{H}_\mathrm{B|A}-H_\mathrm{B|A}.
\end{equation}
The formula for the entropic bound is
\begin{equation}
\label{bound_formula}
B = -2\log_2 C,
\end{equation}
where $C = \max_{jk}||\Pi_j^{1/2} \widetilde{\Pi}_k^{1/2}||_\infty$, with $||A||_\infty$ denoting the largest singular value of the operator $A$ \cite{krishna2002entropic}. This bound is how the complementarity of the temporal/spectral measurement bases underpins the security of the key. As shown in Appendix~\ref{app_ent}, $C$ is well approximated by the quantity $\sqrt{\delta\omega \delta t/2\pi}$, so that finally the size of the secret key is
\begin{equation}
\label{Final_bound}
H_\mathrm{B} \geq I \geq \mathrm{min}\left\{H_\mathrm{B}, \log_2\left[\frac{2\pi}{\delta \omega \delta t}\right]-\widetilde{H}_\mathrm{B|A}-H_\mathrm{B|A} \right\},
\end{equation}
where we have included the natural constraint that the secret key is upper bounded by the marginal entropy $H_\mathrm{B}$ of the measurements (since $H_\mathrm{B|E}\leq H_\mathrm{B}$). That is, regardless of the measurement precision, the secret key cannot exceed the entropy of Bob's marginal statistics. This is in turn bounded from above by $H_\mathrm{B}\leq I_M = \log_2 M$, which is the maximum information content for $M$-outcome measurements. By design $M=(\beta_+/2\beta_-) K$ is roughly equal to the Schmidt number of the source, to within a factor on the order of unity. With appropriate choices of $\beta_\pm$ and for sufficiently resolving measurements (such that $\delta\omega \delta t \sim M^{-1}$), this protocol can therefore be efficient in terms of the information extracted from the quantum states it consumes as a resource. For the modulation scheme described above, we have $\delta t = \delta \omega/\ddot{\phi}$, which along with Eq.~(\ref{phi_fix}) indeed yields $\delta\omega \delta t = (\beta_+\beta_- M)^{-1}$. However we note that Eq.~(\ref{Final_bound}) is a general expression for the size of the secret key for a TFQKD protocol with arbitrary finite resolution in both the time and frequency bases, which is not specific to our proposed modulation scheme and makes no assumptions about the form of the state produced by the source, or the power of Eve's interventions.
\section{Practical considerations}
So far we have considered the security of finite-resolution measurements on photon-pair states, but the real photonic states emitted by a parametric source contain dominant contributions from vacuum (no photons emitted), and smaller contributions from multi-pair emission (most significantly, four photons emitted instead of two). Multi-pair emission does not necessarily represent a security threat, since in general the spectral correlations of the source are strictly pair-wise, meaning that Eve cannot glean extra information by `splitting off' extraneous photons. However our security proof is based on a representation of the POVMs in the Hilbert space spanned by a single photon impinging on each detector. The proof is therefore only approximate, but \emph{squashing protocols} have been proposed for qubit-based QKD to deal with the effects of multi-pair emission \cite{Beaudry:2008uq}, and a similar idea could be used here by randomly assigning measurement outcomes whenever more than one detector in an array fires.

Alice and Bob can remove the influence of the vacuum components by post-selecting on joint detection events, and this also allows the protocol to run when the channel transmission and detector efficiencies are less than unity. For Alice and Bob to use postselection, they must employ spectral filters and temporal gates that prevent signals entering their detectors with spectral or temporal components outside the range of their measurements. To see why, suppose that Eve is capable of performing a very precise projective (non-destructive) frequency measurement on Bob's photon. If he goes on to measure in the frequency basis, he will get the same result as Eve, and if Alice also used the frequency basis, this measurement result will form part of the sifted key. However, if Bob makes a temporal measurement without any time gate, the narrowband photon produced by Eve's measurement will fall mostly outside of Bob's detection range in the time domain, and so will not be detected. Alice and Bob therefore discount this event, since they are post-selecting events where both parties receive a photon. \emph{Mutatis mutandis} Eve can make the same kind of intercept-resend attack in the temporal basis. In this way, Eve can gain perfect knowledge of the sifted key, while forcing Alice and Bob to ignore those events where she failed to choose the correct measurement basis. Alice and Bob can prevent this attack by always rejecting photons occupying regions of chronocyclic space outside the sensitivity regions of their detectors, by means of spectral filters and temporal gates \cite{Qi:2011bh}. In our proposed protocol this could be accomplished with a combination of a dielectric stack and a Pockels cell preceding each of Alice's and Bob's detection systems. See \cite{Ray:2013uq} for a more general discussion of the effects of finite detection range on entanglement verification.

\begin{figure}[h]
\begin{center}
\includegraphics[width = \textwidth]{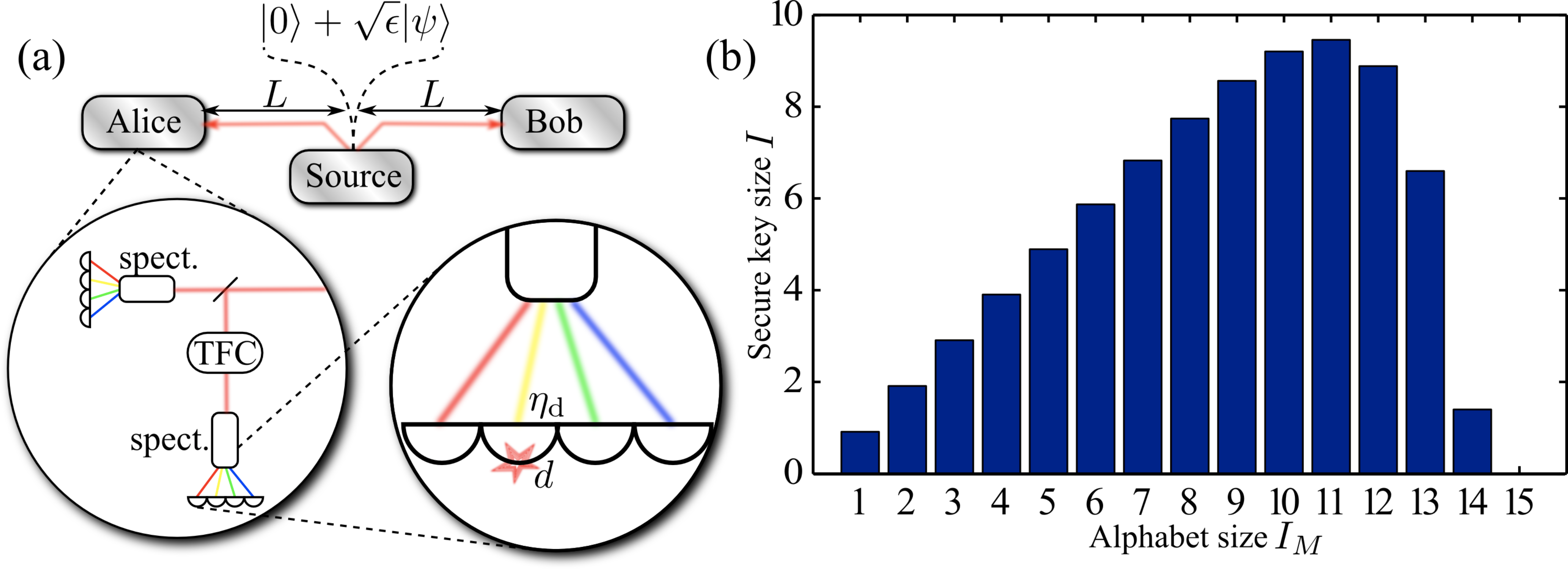}
\caption{TFQKD in the presence of losses and noise. (a) The source emits mostly the vacuum state $\ket{0}$, with $\epsilon$ the small probability to emit a correlated photon pair. For simplicity we consider the source located equidistant from Alice and Bob, connected by lossy channels of length $L$ with attenuation length $L_\mathrm{att}$. The photon detectors comprising the spectrometers have efficiency $\eta_\mathrm{d}$ and suffer dark counts with probability $d$. (b) We compute the secure key size $I$ as a function of the alphabet size $I_M$ for a typical protocol with $\epsilon=0.1$, $\eta_\mathrm{d}=25\%$ (including coupling losses) and $d=10^{-6}$, at a distance $L=L_\mathrm{att}$. Dark counts cause a sharp drop in the secure key size for alphabets larger than 11 bits.}
\label{fig:TF_QKD_dark_counts}
\end{center}
\end{figure}

With postselection as described above, non-unit efficiencies and losses do not directly affect the security, but they reduce the key exchange rate which increases the fraction of uncorrelated dark counts. As is the case with all other QKD protocols, the increasingly important role of dark counts as channel losses grow ultimately limits the distance over which TFQKD can be run securely \cite{ursin-2007-3,Mower:2012cr,Zhang2008}. To analyse this limitation we consider a simple error model in which the source joint spectrum produces perfect correlations at Alice's and Bob's detectors, while random errors occur with probability $p$ in either the time or frequency measurements, so that
\begin{equation}
\label{random_errors}
p^\mathrm{BA}_{ba}= \left[ (1-p')\delta_{ba} + p'/M \right]/M,
\end{equation}
where $p'=Mp/(M-1)$. With a dark count probability $d$ for each of the $M$ detectors in any given run of the photon source, the error probability is found to be (see Appendix~\ref{app_dark}),
\begin{equation}
\label{error_prob}
p = \frac{\kappa(M-1)}{\kappa M+1};\qquad \textrm{where}\qquad \kappa = \frac{2d(1-\eta)}{\eta}+ Md^2\left(1+\frac{1-\epsilon}{\epsilon \eta^2}\right).
\end{equation}
Here $\eta=\eta_\mathrm{chan} \eta_\mathrm{d}$, with $\eta_\mathrm{chan}=e^{-L/L_\mathrm{att}}$ the transmission through a quantum channel of length $L$ and attenuation length $L_\mathrm{att}$ and $\eta_\mathrm{d}$ the overall detector efficiency, assuming Alice and Bob have identical equipment. We have also introduced the photon-pair emission probability $\epsilon$ such that the quantum state emitted by the source is approximately given by $\ket{\!0}+\sqrt{\epsilon} \ket{\!\psi}$, with $\ket{\!0}$ the vacuum state and $\ket{\!\psi}$ given in Eq.~(\ref{psi}) (see part~(a) of Fig.~\ref{fig:TF_QKD_dark_counts}). Note that in a typical implementation with $M\gg 1$ and $\{\eta,\epsilon\}\ll 1$, with a small dark count probability $d \ll \eta \epsilon /M$, we have $p'\approx p \approx 2Md/\eta$. With the error model in Eq.~(\ref{random_errors}) the conditional entropies are both given by
\begin{equation}
\label{error_entropies}
\widetilde{H}_\mathrm{B|A}=H_\mathrm{B|A} = p\log_2(M-1)+h(p),
\end{equation}
where $h(x)=-x \log_2 x - (1-x)\log_2 (1-x)$ is the binary entropy. Substituting these expressions into Eq.~(\ref{Final_bound}), using $\delta\omega\delta t = 1/\beta_+\beta_-M$, one obtains
\begin{equation}
\label{approxI}
I \geq I_M-2p\log_2(M-1)-2h(p)-c,
\end{equation}
where we have defined the `binning deficit' $c=-\log_2(2\pi\beta_+\beta_-)$. The security condition $I\geq 0$ is of the same form as that derived in \cite{Bourennane:2002fk}, which considered QKD with finite dimensional quantum systems \cite{Cerf2002}\footnote{Eq.~(27) in \cite{Bourennane:2002fk} corresponds to setting $I=0$ with $c\longrightarrow0$, $M\longrightarrow N$ and $p\longrightarrow e^N_B$.}. Therefore we see that TFQKD, based on the continuous chronocyclic degree of freedom, can be treated as a finite-dimensional QKD protocol, with a small correction factor represented by $c$, that takes account of the choices made in binning the measurements.

A realistic joint spectrum, as approximated in Eq.~(\ref{JSA}), will not exactly produce the simplified delta-correlated outcome distribution in Eq.~(\ref{random_errors}); in general there will be some finite cross-talk between adjacent detection channels and a Gaussian-like variation in the marginal detection probabilities. However we have verified numerically, using Eq.~(\ref{JSA}), that the choices $\beta_+ = 3/4$, $\beta_- = 1/5$, which we introduced previously, generate $H_\mathrm{B}\approx I_M$, and agree with Eq.~(\ref{error_entropies}), to within 0.05~bits, so that Eq.~(\ref{approxI}) is approximately correct even for a source with a realistic joint spectral amplitude.

Part~(b) of Fig.~\ref{fig:TF_QKD_dark_counts} shows the variation of the secure key size $I$ on the size $I_M$ of the alphabet for a typical implementation with APD-type detectors and a channel length with $L=L_\mathrm{att}$ (typically $\sim$20~km in the C-band), assuming the values of $\beta_\pm$ used previously. The secure key initially grows in proportion to the alphabet, albeit with a magnitude reduced by the binning deficit $c\approx 0.086$ bits. Here the deficit is small enough that exchanging up to 4 secure key bits per detected photon pair remains feasible with the components described previously. However for $I_\mathrm{M}>11$ (\emph{i.e.} $M>2048$) the secure key decreases and rapidly falls to zero, since the dark count rate from the growing number of detectors causes significant errors that are attributed to Eve. This calculation exemplifies a natural technical trade-off in TFQKD, where information can be extracted rapidly from photon pairs using detector arrays as we propose here, whereas single-detector protocols extract less information per registered photon pair, but are affected much less by dark counts \cite{Mower:2012cr}.
\section{Conclusion}
We have described an implementation of highly multiplexed time-frequency quantum key distribution using photon-counting spectrometers and phase modulators to implement frequency and arrival-time measurements on pairs of spectrally correlated photons by means of time-to-frequency conversion. We analysed the technical demands required for such a protocol and found that a large-alphabet detection system is feasible with commercially-available components. We developed a general security analysis for continuous-variable QKD protocols with binned measurement results, and applied it to our protocol to model the effects of dark counts on the size of the secure key. The protocol we presented is one of a family of time-frequency protocols well suited to transmission using waveguides and optical fibres, where spatial or polarisation mode encoding is vulnerable to cross-talk and differential scattering losses. Highly multimode quantum light sources are now a well-developed technology, and the use of entangled beams allows the communication distance to be extended via entanglement swapping with quantum repeaters. Our measurement scheme based on phase modulation represents a compact and technically straightforward way to demultiplex a large-alphabet key from a multimode entangled light source, which is a convenient source of correlations that removes the need for an active encoding step. The security analysis we presented shows how the protocol can be treated as a finite-dimensional QKD scheme, despite the continuous nature of the chronocyclic degree of freedom. We anticipate that these results will influence the next generation of quantum cryptographic systems.

\appendix
\section{Entropic uncertainty relation for finite-resolution measurements}
\label{app_ent}
\begin{figure}[h]
\begin{center}
\includegraphics[width = \textwidth]{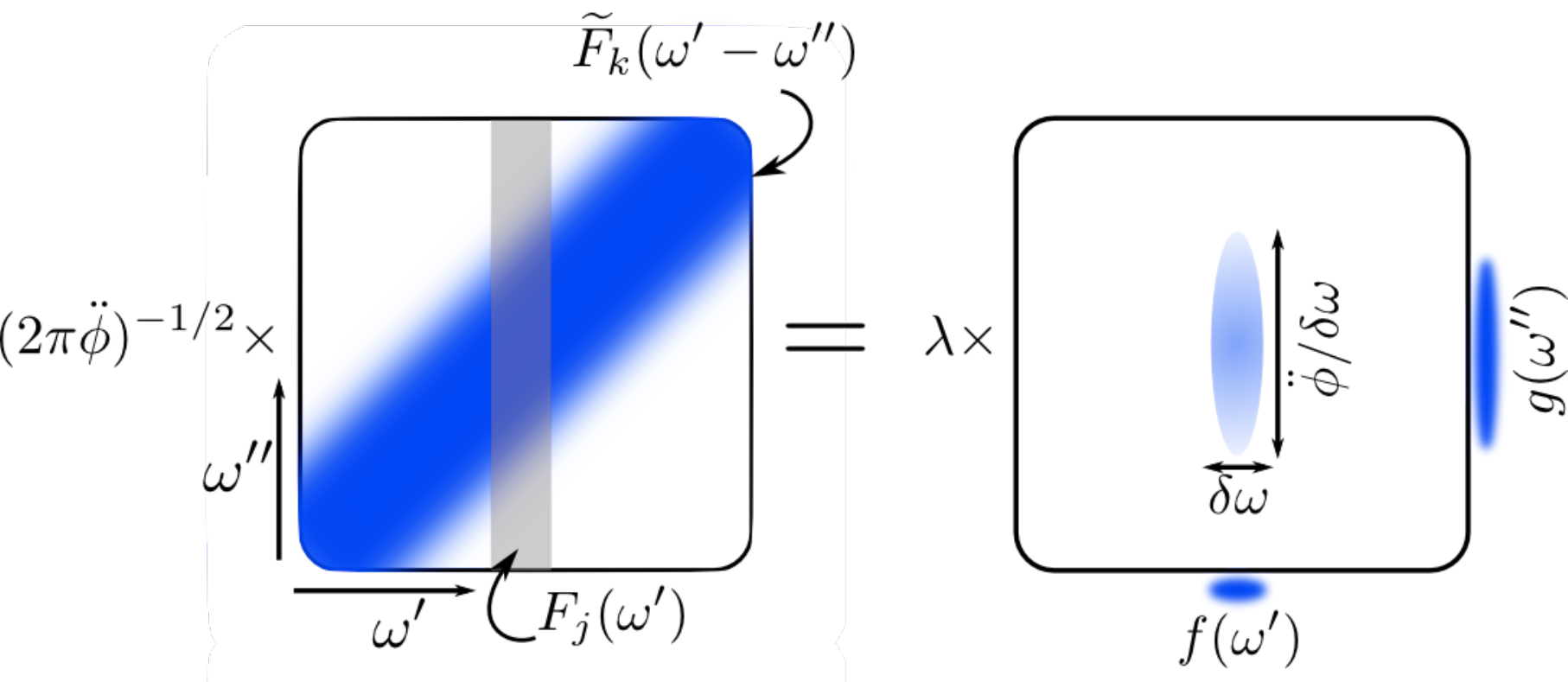}
\caption{Schematic of the operator $\Pi_j^{1/2} \widetilde{\Pi}_k^{1/2}$, showing the approximate computation of its largest singular value.}
\label{fig:appendix_fig}
\end{center}
\end{figure}

As described in the main text, the complementarity of non-commuting measurements $\{\Pi_j\},\{\widetilde{\Pi}_k\}$ is manifested by the entropic uncertainty relation in Eq.~(\ref{entropic_bound1}), with the bound $B = -2\log_2 \max_{jk}||\Pi_j^{1/2} \widetilde{\Pi}_k^{1/2}||_\infty$. To compute the singular values of the product $\Pi_j^{1/2} \widetilde{\Pi}_k^{1/2}$ we first note that the square-roots can be dropped because for the top-hat-shaped frequency bins defined in Eq.~(\ref{F_bin}) we have $F(\omega)^{1/2}=F(\omega)$. We therefore consider the singular values of the operator
\begin{eqnarray}
\label{sing_val_1}
\Pi_j \widetilde{\Pi}_k &=&(2\pi \ddot{\phi})^{-1/2}\int \!\!\! \int F_j(\omega')\widetilde{F}_k(\omega'-\omega'')\ket{\!\omega'}\bra{\omega''\!}\rmd\omega'\rmd\omega''\\
\nonumber &=& \frac{\delta\omega}{2\pi \ddot{\phi}} \int \!\!\! \int F_j(\omega')e^{-\mi \omega_k (\omega'-\omega'')/\ddot{\phi}}  \mathrm{sinc}\left[ \frac{\delta\omega}{2} \frac{(\omega'-\omega'')}{\ddot{\phi}}\right]\ket{\!\omega'}\bra{\omega''\!}\rmd\omega'\rmd\omega''.
\end{eqnarray}
The phase factor $e^{-\mi \omega_k (\omega'-\omega'')/\ddot{\phi}}$ can be dropped since this represents a unitary transformation that does not affect the singular values. Similarly the index $j$ on $F_j$ is arbitrary since this parameterises a shift in the position of the top-hat function that again does not affect the singular values. The singular values are thus obtained by considering the Schmidt decomposition of the two-dimensional function
\begin{equation}
\label{2_dim_fun}
\mathcal{F}(\omega',\omega'')=
\left\{\begin{array}{cc}\frac{\delta\omega}{2\pi \ddot{\phi}}\mathrm{sinc}\left[ \frac{\delta\omega}{2} \frac{(\omega'-\omega'')}{\ddot{\phi}}\right]\quad & | \omega'-\omega_j | \leq \delta\omega/2;\\
0 \quad& \quad\quad\quad\;\mathrm{otherwise}.\end{array}\right.
\end{equation}
As shown in Fig.~(\ref{fig:appendix_fig}), this has the form of a narrow vertical strip cut from a broad diagonal band, and is well approximated by the factorable function $\mathcal{F}(\omega',\omega'')\approx \lambda f(\omega')g(\omega'')$, where $f(\omega')$ is a mode function with width $\delta \omega$, $g(\omega'')$ is a mode function with width $2\pi \ddot{\phi} / \delta \omega $, and
\begin{equation}
\label{lam}
\lambda = \frac{\delta\omega}{2\pi \ddot{\phi}}\times \sqrt{\delta \omega \frac{2\pi \ddot{\phi}}{\delta \omega}}.
\end{equation}
Numerical calculations confirm that this remains an excellent approximation to the largest singular value of the function in Eq.~(\ref{2_dim_fun}) provided $\delta \omega / ( 2\pi \ddot{\phi} / \delta \omega) < 0.1$, which is guaranteed by Eq.~(\ref{phi_fix}) for a large-alphabet protocol with $M \gg 1/\beta_+\beta_-$. Substituting Eq.~(\ref{lam}) into Eq.~(\ref{bound_formula}) and using $\delta t = \delta \omega / \ddot{\phi}$, we obtain the bound on the secret key given in Eq.~(\ref{Final_bound}). We note that the resulting bound applies quite generally to any TFQKD protocol employing binned frequency and time measurements with resolutions $\delta\omega$, $\delta t$.
\section{Dark count error probability}
\label{app_dark}
The error probability $p$ is given by \cite{Bourennane:2002fk}
$$
p = \frac{P_\mathrm{incorrect}}{P_\mathrm{correct}+P_\mathrm{incorrect}},
$$
where $P_\mathrm{correct}$, $P_\mathrm{incorrect}$ are the probabilities for recording correlated and uncorrelated detector clicks, respectively. Alice and Bob postselect on single clicks at each side, so only events where one detector fires each are counted. We make the simplifying assumption that if both photons from a photon pair are successfully emitted and detected, without a dark count occurring in any other channel, then the two registered outcomes are perfectly correlated. If these photons are not registered, but dark counts occur, it is possible that the outcomes are incorrectly correlated. Taking into account all possibilities yields
\begin{equation}
\label{Pinc}
P_\mathrm{incorrect} = \overline{d}^{2(M-1)}[2 \epsilon \eta \overline{\eta} (M-1) d + \epsilon \overline{\eta}^2 d^2 M(M-1)+\overline{\epsilon} d^2 M(M-1)],
\end{equation}
and
\begin{equation}
\label{Pcorr}
P_\mathrm{correct} = \overline{d}^{2(M-1)}[\epsilon \eta^2 + 2\epsilon \eta \overline{\eta} d + \epsilon \overline{\eta}^2 d^2 M +\overline{\epsilon} d^2 M],
\end{equation}
where the overbar notation $\overline{x}=1-x$ denotes the probabilistic complement. The common factor of $\overline{d}^{2(M-1)}$ represents the probability that there is no dark count in the $M-1$ remaining bins on each side, and this factor cancels out in the expression for $p$. After some algebra, one obtains the expressions in Eq.~(\ref{error_prob}).

\section*{Acknowledgements}
The authors gratefully acknowledge helpful discussions with M.~G.~Raymer, S.~J.~van~Enk and N.~K.~Langford in the initial stages of this work. BJS was supported by the University of Oxford John Fell Fund and EPSRC grant No.~EP/E036066/1.

\end{document}